# Fog Computing for Detecting Vehicular Congestion, An Internet of Vehicles based Approach: A review

Arnav Thakur and Reza Malekian, Senior *Member, IEEE*

*Abstract*—Vehicular congestion is directly impacting the efficiency of the transport sector. A wireless sensor network for vehicular clients is used in Internet of Vehicles based solutions for traffic management applications. It was found that vehicular congestion detection by using Internet of Vehicles based connected vehicles technology are practically feasible for congestion handling. It was found that by using Fog Computing based principles in the vehicular wireless sensor network, communication in the system can be improved to support larger number of nodes without impacting performance. In this paper, connected vehicles technology based vehicular congestion identification techniques are studied. Computing paradigms that can be used for the vehicular network are studied to develop a practically feasible vehicular congestion detection system that performs accurately for a large coverage area and multiple scenarios. The designed system is expected to detect congestion to meet traffic management goals that are of primary importance in intelligent transportation systems.

## I. Introduction

VEHICUALR congestion is a global phenomenon. It is impacting the efficiency of the transport sector by introducing delays which cause increases in travel time and leads to increased fuel consumption and associated vehicular emissions [1], [2], [3]. Vehicular congestion causes loss in productivity and the monetary loss is estimated to be as high as 160 billion dollars per year in the U.S alone [3]. Of the total cost, 85 % of the economic cost of congestion is due to loss of work hours caused by increased travel time, 13 % of the loss is attributed to fuel wastage and increased vehicular emissions is responsible for the remaining 2 % of the loss [2].

Vehicular congestion is expected to severely worsen in magnitude due to rapid increase in the number of vehicles. In a study conducted by IBM, it was estimated that current global vehicular population is over a billion and is expected to double by the year 2020 [3]. To enhance the efficiency of the transportation sector and reduce the associated economic and environmental consequences, an accurate real-time vehicular traffic congestion identification system is required to be used in effective traffic management schemes that improve traffic flow and utilization of road infrastructure.

Vehicular congestion is quantified by using the traffic density estimate metric which is the number of vehicles per unit area [4]. The traffic density estimation techniques use parameters that include road occupancy rates, difference of incoming and outgoing traffic in an observation area, traffic flow estimation and motion detection and tracking to identify congestion [5].

Vehicular congestion detection using traffic density estimation is performed by using contact and non-contact based sensing techniques [4], [5], [6], [7]. In contact based techniques, the number of vehicles passing the sensory node are detected by loop detectors and pressure pads [8], [9]. Contact based methods are not reliable as accuracy is dependent on maintenance of sensing apparatus and results are valid for a small coverage area [5], [8]. Thus, contact based techniques are not suitable for estimating traffic density in a large observation area as it would incur high deployment and maintenance costs of sensory infrastructure.

Traffic density estimation for detecting vehicular congestion using non-contact based sensing techniques deploy surveillance camera, microphones and connected vehicle technology (CVT) [4], [5], [7], [8]. Surveillance cameras based non-contact techniques capture the traffic stream. Segmentation and motion tracking algorithms are applied on vehicle objects in the video sequence to extract motion related parameters which are processed by classifiers to determine congestion levels [10]. In microphone based sensing technique, cumulative acoustic signals are acquired that contain components of engine noise, idle engine noise and tire noise [11]. Traffic congestion is determined based on the proportion of the above mentioned components in the input signal [11].

Non-contact traffic density estimation for congestion detection based on connected vehicle technology use the Internet of Vehicles (IoV) approach. IoV are Internet of Things (IoT) based solutions for vehicular clients designed to achieve the road safety and traffic management goals of intelligent transportation systems (ITS) [12], [13], [14].

A wireless sensor network (WSN) is used in the CVT based

This work was supported in part by the National Research Foundation, South Africa under Grant IFR160118156967 and RDYR160404161474.

Arnav Thakur is with the Department of Electrical, Electronic and Computer Engineering, University of Pretoria, South Africa.
Reza Malekian is with the Department of Electrical, Electronic and Computer Engineering, University of Pretoria, South Africa. Email: reza.malekian@ieee.org

IoV approach. The WSN contains vehicular nodes which function as sensory nodes. Stationary waypoints and infrastructure related nodes function as roadside units (RSUs) which is the link with the network infrastructure and a handler node operating as the centralized server [14], [15], [16]. The WSN uses vehicle to vehicle (V2V) communication for exchanging data between vehicular sensory interfaces and vehicle to infrastructure communication (V2I) for exchange of data between vehicular nodes and RSUs [13], [17]. Parameters collected from the vehicular sensory nodes include vehicle position, direction and route, frequency of braking events, surface area of vehicles and average velocity and acceleration of vehicle node [2], [3], [18]. The collected parameters are processed by the traffic density estimation algorithms for identifying congestion on the processing node which varies across IoV architectures being used [4], [6], [7], [19].

Results from non-contact based techniques such as microphone and surveillance cameras are not reliable as the results are dependent on environmental conditions [4], [11]. Moreover, results from surveillance cameras are not in real-time due to computationally intensive nature of the image processing algorithms used for segmentation and tracking of traffic stream [10].

Results obtained from CVT based technique are not affected by weather and environmental conditions. Additionally, the sensory data is ready for processing in real time. Mobile nature of the sensory nodes used in CVT based approach enables this technique to be used for large deployment areas with low costs and higher reliability of the inputs [4], [20]. Thus, CVT based techniques for congestion identification are reliable and are suitable for usage in an effective traffic management system [2], [19], [21].

Vehicular networks require collection and processing of large amounts of data in real-time. The information being exchanged is time-sensitive due to high mobility of the nodes and the communication links have limited bandwidth. The network is also required to be context location aware [22]. By the year 2020, it is estimated that there will be up to 152 million active connected vehicles that will generate more than 30 TB of data every day [16]. Thus, the communication system has to be robust for a practically feasible solution that meets Quality of Service (QoS) and Quality of Performance (QoP) requirements [15], [16].

Distributed cloud computing based approaches of Fog Computing are used in conjunction with cloud computing to optimize communication in the WSN by reducing the number of data exchange events [23]. Fog computing uses localized computing and lightweight algorithms to process sensor data at edge devices which optimizes the capacity of the communication and computational components of the system [24]. In this study, Fog Computing for IoV systems are studied for developing a practically feasible solution for detecting vehicular congestion.

Remainder of this paper is organized as follows: In section II, multiple cloud architectures suitable for WSNs are studied and compared. In section III, vehicular Fog based cloud computing architecture for IoV applications is studied. In section IV vehicular congestion detection using CVT based WSN and Fog Computing based principles are studied. Results from the studied literature is discussed in section V and a conclusion is drawn in section VI.

## II. Vehicular Computing Models

The wireless sensor network in an IoV system acquire physical world data from the underutilized onboard sensors of the vehicles which function as mobile probes for applications such as Advanced Vehicle Control and Safety Systems (AVCSS), Advanced Traffic Management Systems (ATMS), Advanced Traveler Information Systems (ATIS), Advanced Public Transportation. Systems (APTS) ,Commercial Vehicle Operation Systems (CVOS), urban surveillance and environmental monitoring [16], [18], [25], [26], [27], [28]. The WSN shares storage, computational and communication resources to maximize utilization and effectively perform the processing of the large voluminous sensory information for the desired application [29]. Components of the vehicular WSN are summarized in Fig 1.

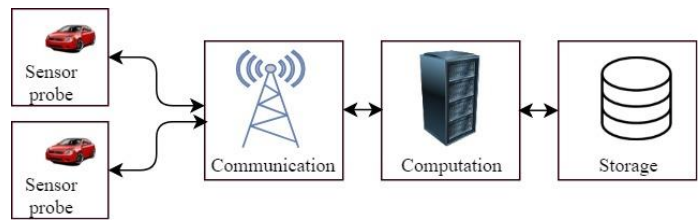

fig. 1. Components of the vehicular WSN in IoV cloud based approach which are shared to maximize utilization and increase responsiveness for the application ([29]).

Vehicles which function as the mobile probes contain sensors in a package referred as the onboard unit (OBU) [13]. The OBU contains the communication apparatus. Communication interfaces include inter vehicle communication channel for V2V communication and V2I communication for connectivity between the vehicle and RSU with remote handler [14].

The sensors included in the OBU include safety related distance and night vision sensors, kinetics related sensors for tracking motion related parameters such as speed and acceleration and positioning apparatus such as GPS for monitoring applications [18]. The OBU also contains storage and computing resources to process sensory information locally [16].

The WSN network architecture aims to optimally utilize the communication and computational resources by collaborative sharing and the cloud computing services types realized in vehicular networks include Network as a service (NaaS), Cooperation as a service (Coaas), Computation as a service (COaaS), storage as a service (STaaS), sensing as a service (SEaaS) [13], [29]. NaaS is used to facilitate communication links between the nodes and the remote handler, SEaaS is used for real time monitoring of physical world related parameters, COaaS and STaaS is used for optimally exploiting the underutilized computational capability of the vehicular probes and CaaS is used for mutual collaboration between vehicular nodes [13], [29]. Cloud based services for multiple applications in the IoV based system are handled through web sockets, Restful and Java APIs and protocols that include Constrained Application Protocol (CoAP), Message Queue Telemetry Transport (MQTT) and Advanced Message Queuing protocol (AMQP) [13], [30]. Architecture of the vehicular cloud in IoV based approach is summarized in Fig 2.

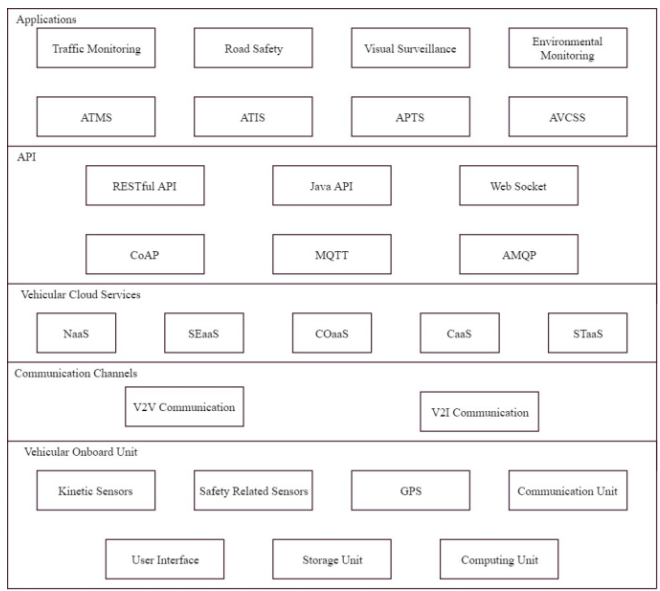

Fig. 2. Architecture of the vehicular WSN in IoV cloud based approach ([13]).

Multiple computing paradigms exist to realize vehicular sensor networks. This includes vehicular cloud computing (VCC), vehicles using cloud (VuC), mobile cloud computing (MCC) and vehicular Fog Computing (VFC). In the following, computing paradigms are studied for application in vehicular networks for IoV based solutions and a comparison is summarized in table I.

*A. Vehicle using Cloud (VuC)*

In the Vehicle using Cloud (VuC) approach, a group of connected client vehicles avail the services of the conventional cloud through the internet [29]. This enables a vehicle to access a wide pool of configurable processing and storage resources in the data centre which are useful for applications such as real time traffic monitoring [13], [29]. VuC based systems are least prone to limitations of storage and processing resources as they are shared between all clients [13]. VuC based networks cannot be formed autonomously between vehicles for mutual collaboration for processing of sensory data [13]. VuC is prone to large communication delays due to limited bandwidth related constraints of the data link with the remote data centre and makes it unsuitable for deployment in applications with highly mobile nodes such as for vehicular clients [15].

*B. Mobile Cloud Computing (MCC)*

In Mobile Cloud Computing (MCC) based systems storage and computation operations of mobile nodes are outsourced to other entities [22]. Storage and computations are performed on lightweight cloudlet servers placed on the edge of the network [23].

MCC enables other mobile nodes to access and benefit from the information related to the stored data and computations being performed by a central entity such as a RSU for vehicular applications [13]. Low cost solutions can be developed by using MCC which increase utilization of shared resources and have nodes with low power consumption, storage and computational capability [13], [15].

MCC enables mobile devices to access powerful computational resources, but is not practical for real-time applications as it involves time consuming uploading of data to the handler on a client-server connection and the reliability is dependent on the performance of the communication infrastructure [15].

*C. Vehicular Cloud Computing(VCC)*

Vehicular Could Computing based networks contain an autonomous group of connected vehicles that from a cloud to dynamically share their sensing, communication, storage and computation resources [15], [29], [31]. The shared resources are made available in areas with high concentration of vehicular nodes such as traffic intersections and parking lots. The flexible and dynamic resources are shared between vehicles by using peer to peer and client server connections over inter vehicle communication interfaces [13]. The approach aims to utilize the increasing computational and sensing capability of onboard computers in vehicles which are getting increasingly sophisticated.

Autonomous and self-organized nature of the network makes it ideal for mobile environments based on traffic distribution [15]. This enables VCC based systems to resolve unexpected events on demand in real-time and makes it well suited for applications such as congestion detection and traffic management [31]. Additionally, stationary vehicles at major parking lots such as at airports and shopping malls can be utilized as data centers for processing large scale vehicular sensory data [31].

The storage and computation resources in VCC is independent of traffic distribution [13]. Additionally large number of vehicles in the networks can increase latencies and influence reliability of the system as the communication link's bandwidth is also a constraint [13], [15].

*D. Fog Computing (FC)*

Edge sensory devices lack adequate storage, bandwidth, computational capability and are powered by a battery based power source. IoT based systems depend heavily on powerful server infrastructure contained in the cloud to provide large elastic resources [22]. The solution introduces large round trip times for communication between the sensory nodes and the server and deems it unsuitable for deployment in time sensitive applications such as for vehicular networks for transportation related applications [15], [22]. The centralized approach is unsuitable for systems with large geo distribution of sensory nodes and high mobility [23]. Thus, a scalable solution is required that does not negatively influence the responsiveness and reliability of the sensory nodes that operate on the network end points.

Fog computing based approach is evolving with the increase in the number of IoT devices and services and is used in conjunction with traditional cloud computing which was first proposed by Cisco in 2012 [23]. Fog computing is a distributed computing based approach where the Fog layer is an intermediate layer in the three-layer framework. The first layer in a Fog system contains edge devices equipped with sensors and raw processing capability of the collected data [24], [32]. The Fog computing level nodes process and store data to make neighborhood level decisions and the cloud computing level is used for wide area level advanced decision making [32], [33].

The Fog layer is used in-between the IoT sensors and cloud servers which functions as a local cloud for nodes at the edge of the network [22], [23]. This is done to bring elastic resources such as computations and storage provided by the traditional cloud closer to the edge devices distributed over a large coverage area with mobility support [22].

Having a cloud like infrastructure closer to the edge devices helps to minimize network traffic and amount of information being processed on the central cloud is reduced [33]. Thus, IoT services can be offered to a large number of devices over a wide coverage area with low latency and high bandwidth with high QoS and QoP [23].

Nodes at the Fog layer have limited resources and this directly impacts the fault tolerance capability of the Fog layer nodes [34]. Additionally short range communication technologies such as Wi-Fi and ultra-wideband (UWB) are used for exchanging data between the Fog Nodes and edge devices which are prone to interference [34]. This challenge can be addressed for vehicular applications by using Dedicated Short Range Communication (DSRC) based wireless communication technology which is less prone to interference [35].

The decentralized approach of the distributed Fog computing approach makes it suitable for a wide range of applications ranging from healthcare to smart grid, augmented reality and vehicular networks [33].Various computing models studied in this section are compared in table I.

TABLE I
COMPARISON OF CHARACTERISTICS OF COMPUTING MODELS ([13], [15], [22], [33])

| Characteristic | VuC | MCC | VCC | FC |
|---|---|---|---|---|
| Storage resources | Unlimited | Low | Medium sized | High |
| Computational capability | Unlimited | Low | Medium size | High |
| Architecture | Centralized | Centralized | Autonomous | Distributed |
| Mobility support | Limited | Yes | Yes | Yes |
| Response time | High | High | Low | Low |

From the above compared computing models, VuC has unlimited storage and processing capability as it is backed by configurable resources. Fog Computing based systems has high storage and processing capability provided by the fog and cloud layer. VCC based systems have lower storage and processing capability when compared to FC and VuC. MCC based systems have the lowest storage and computational capability.

VuC based systems with large storage and computational resources and centralized architecture have the lowest support for mobile nodes and have high response time. This makes VuC based system unsuitable for vehicular traffic detection applications.

Whereas, MCC has a centralized architecture with support for mobile nodes. MCC suffers from high response times and limited storage and computational capability. This makes MCC unsuitable for applications for vehicular traffic detection.

Mobility support along with low response time in VCC and FC based systems make them suitable for usage in vehicular congestion detection applications. Fog computing based systems have higher storage and processing capabilities. The optimized communication and usage of lightweight algorithms in Fog computing give it an edge over VCC. This makes Fog computing best suited for vehicular congestion detection applications which require scalable solutions with fast processing.

### III. VEHICULAR CLOUD AND FOG COMPUTING

The system in a connected vehicle technology based approach with Fog computing contains three layers which are the data generation, fog and cloud layer [16]. The edge devices part of the data generation layer gather sensory data and consume IoV services are the OBUs placed in the vehicles [3] [18]. Sensors used in the mobile probes at the network edge include speed sensors, distance sensors, radars, accelerometer, gyroscope, GPS, tilt sensors and image and audio sensors for surface analysis [4], [28], [36], [37]. Sensory information from the vehicle's ECU is also acquired from the OBD II interface [27]. Sensory data is preprocessed to make decisions applicable for the vehicle that gathers the sensory information.

The Fog nodes in the IoV system are the stationary waypoints contained in the RSU and act like a bridge between the data generation layer contained in the vehicle sensor probes and the cloud. Data is exchanged using V2V and V2I communication which use long and short range wireless communication technologies such as ZigBee, Wi-Fi, Dedicated Short Range Communication (DSRC) and Wireless Access for Vehicular Environments (WAVE) [30], [36]. Data flow in the entire environment is real time or event driven which influences the availability of resources and services in the system [23]. A publish/subscribe based model using Fog Computing for IoV systems is proposed in [38]. Fog nodes collaborate using techniques that include cluster based on the homogeneity of the nodes, master-slave and peer to peer (P2P) [23]. The Fog nodes process the data sent by the vehicles to make local area based decisions and report processed data to cloud servers. [16].

The cloud based servers perform computationally intensive operations on the information sent by the RSUs to make holistic decisions such as managing and controlling of road infrastructure for a large area and provide centralized control from a remote location [16]. The architecture used in Fog Computing for vehicular networks is summarized in Fig 3.

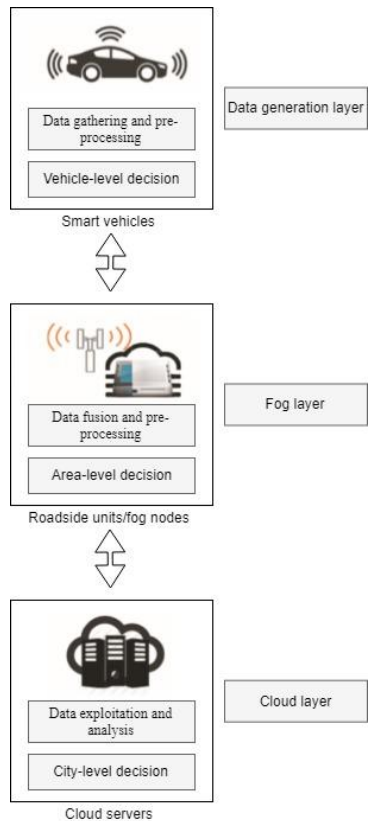

Fig. 3. Three layer architecture used in Fog Computing for vehicular networks (From [16]).

In [2], a traffic management system using Fog Computing principles is developed where the entire city is divided into regions and independent RSU units with storage, processing and wireless communication capabilities functioning as Fog nodes are responsible for detecting and controlling congestion in its area of interest in the region. The size of the area of interest is limited by the range of the radio in the RSU and vehicles exchange sensory information with the nearest RSU [2]. A similar architecture is used in [3] for a route management mechanism based traffic control system.

## IV. VEHICULAR NETWORKS BASED VEHICULAR CONGESTION DETECTION

Vehicular congestion is identified by observing the average speed of the traffic stream and the traffic density metric, which is the number of vehicles per unit area of length such as vehicles/cell and vehicle/km/lane [4]. The average speed and traffic density related information is used by classifiers to group traffic into various fuzzy sets and is also used to identify congestion levels ranging from free flow, slight to severe [39]. Classifiers used to identify congestion levels include Naive Bayes, K-nearest neighbor, learning vector quantization (LVQ) and support vector machine (SWM) [4], [5], [40].

Driving patterns of motorists are highly unpredictable which depends on multiple parameters and varies for every situation and traffic condition [41]. This makes neural networks based machine learning tools ineffective for congestion detection and handling. This is however not true for autonomous vehicles as reaction to external events can be predicted accurately and thus machine learning techniques can be used for scenarios with autonomous vehicles [41].

Traffic density is estimated in connected vehicle technology by applying algorithms in the RSU on the preprocessed sensory data collected from the vehicular sensory probes by using the communication channels of the vehicular network in the IoV system. The techniques used in the algorithms to identify congestion by using the traffic density metric include vehicle kinetics analysis, motion detection and tracking, average traffic velocity estimation, road occupancy rate estimation and traffic flow rate estimation [4], [5]. In this section, various traffic density estimation techniques using connected vehicle technology are studied for identification of vehicular congestion.

### A. Estimation using Statistical Methods

Statistical methods based traffic density is performed by selecting a volunteer probe vehicle [42], [43], [40]. Statistical distributions based on the number of vehicles in the neighborhood are applied on the data collected by the probe and is used to determine the maximum likely-hood estimate [4], [7]. The global traffic density estimate can be determined by using the local traffic density estimate [7]. A cluster of vehicles with direct communication links can be used as the probes to improve the estimation. Statistical method based estimation is best suited for scenarios where the inter-vehicle distance follows an exponential distribution [7].

In [44], a window time based histogram model approach is used to develop an unique traffic density estimation technique for urban areas. The technique aims to reduce the amount of sensory data collected to make the estimation. The histogram based approach is adopted to reduce the data set required to be analyzed to observe characteristics that are dynamic in nature of individual lanes. Sensory systems are used to collect density patterns for an observation period segmented into sampling events. The histogram used in this approach is a probability distribution function representing the relation between number of vehicles in an observation window and its corresponding probability. Knuth's rule, an optimized version of the Bayesian fitness function is used in this approach to model. A similar lightweight histogram based approach is also developed in [43], to predict periodically, the level of congestion per lane prior to its occurrence.

In [40], statistical methods are applied on GPS data collected from sensory nodes to detect traffic congestion. The collected GPS data is processed to identify on and off road traffic and average speed is computed for the nodes in the data set by using the Haversine formula. The K-Means algorithm is used to classify the identified clusters based on the centroid of the cluster group. The classified clusters groups are processed using an algorithm based on the Naive Bayes method to identify congestion levels.

The performance of statistical methods based traffic density estimation improves with increase in traffic density. This makes statistical methods based traffic density estimation poor for scenarios with low traffic densities and ideal for urban areas which experience high traffic volumes [4], [10]. The performance of statistical methods based congestion identification can be improved by increasing the sampling size and frequency along with increasing the monitoring area covered in a cluster group. This increases the overhead on the communication apparatus and directly impacts the effectiveness of the CVT based technique, especially in scenarios with extremely high traffic volumes [4].

*B. Estimation using V2V Communication*

Neighbor discovery based approach is used in V2V based traffic density estimation technique to mutually exchange sensory information between peers and identify local traffic densities [4]. This limits the coverage area of V2V based traffic density estimation technique to one-hop neighbor [7].

In [4], a three phase approach is developed for estimating traffic density by using V2V communication and one-hop neighbor discovery. The algorithm computes the local traffic density estimate by using one-hop density and average speed of the neighbors which is propagated to the following vehicles in the traffic stream through one-hop broadcasting. In [45], a similar approach is adopted with adaptive broadcasting. The broadcasting event only occurs if the traffic density estimate of the vehicles is higher than its peer which optimizes the number of packets being exchanged over the V2V communication channel.

In [46], traffic density estimation is based on stopping times of vehicles and mobility patterns. The local traffic density is computed by taking the relationship between the number of neighboring vehicles and distance in the front and rear of the vehicle. Accuracy of the estimation by this algorithm is heavily dependent on determining the distance on the front and rear ends of the vehicles.

*C. Estimation using V2I Communication*

Traffic density estimation techniques using V2I based methods collect periodic sensory data by using V2I communication interfaces from multiple probes which are the OBUs mounded on vehicles. Centralized processing is performed on the collected data to estimate the traffic density of the road segment. This gives V2I based techniques much larger coverage area when compared to V2V based techniques [4].

A traffic density function is used in V2I based methods. The function is a polynomial where the coefficients used are derived from topographic feature of the city where the system is to be deployed. The topographical features of the city that are studied to compute the coefficients of the polynomial include average lanes/street, street lengths and street/junction ratio [1], [6]. Vehicles send beacon messages periodically to the RSUs which are handled using queues [47]. The average beacons received from every vehicle is used by the traffic density estimation polynomial [1]. Accuracy of V2I based estimation technique is improved by including additional information in the beacon messages such as speed change and lane change events [4].

V. DISCUSSION

Vehicular congestion detection techniques involve contact based techniques which have an accuracy dependent on maintenance of the sensory infrastructure and non-contact methods. Performance of noncontact based traffic density estimation techniques using surveillance cameras and microphones are influenced by environmental factors and results are not in real-time as they rely on computationally intensive image processing and acoustic processing. It was found that contact based traffic density estimation using embedded loop detectors has an accuracy of 80% and the accuracy obtained with 20% of the traffic stream equipped with connected vehicle technology was 85% [8], [48]. IoV based systems using CVT are best suited for vehicular congestion detection applications that are time sensitive and require high accuracy of sensory data.

Accuracy levels of traffic density estimation suing V2V and V2I based techniques are in acceptable limits. The accuracy levels vary for different scenarios depending on the topographic features of the area being monitored. The error level in estimation of traffic density of -4.86% is achieved using V2V based technique with an estimation polynomial of order 3 [7]. Whereas the error in traffic density estimation by using V2I based methods with a Taylor series density approximation function is 8.34% [7].

The coverage area of the more accurate V2V based techniques is narrow which makes it unsuitable for congestion handling. Whereas V2I based techniques have wide coverage areas and are suitable for congestion handling [4]. Higher accuracy levels of 98 % and wide coverage areas can be achieved by combining V2V and V2I based traffic density estimation techniques [7].

Various computing paradigms exist that can be deployed in the IoV system for congestion detection using CVT. The requirements for IoV systems for congestion detection where large amounts of data are collected for processing include low response time, low communication latency, support for mobility and location awareness. The above requirements are fulfilled by VCC and VFC based approaches. However, VFC based approach uses lightweight algorithms and optimized communication with the servers to improve response time of the system and support a larger number of devices. A three tier Fog Computing based approach was used for a smart home application in [24] and it was found that network traffic can be reduced by 95%. by using Fog Computing principles.

Additionally communication latencies in the Fog Computing based systems is limited to a few tens of milliseconds [34]. This makes vehicular fog computing most suitable for deployment in traffic density estimation applications for congestion detection where data is required to be collected periodically from a large number of vehicular sensory probes to offer a wider coverage area.

Extremely low communication latencies with support for mobility and distributed architecture offered by Fog Computing based approach can be beneficial for applications that require tracking and controlling of vehicles with high accuracy. Deployment of Fog Computing based principles in systems such as vehicular platooning and autonomous control of intelligent vehicles can be investigated in future works [37], [49], [50].

Research on security risks and mitigation of threats in vehicular fog computing is at its initial stages which makes security a challenge as most security mechanisms can only handle threats posed by passive attacks [16].

## VI. Conclusion

Loss in productivity due to the economic and ecological consequences of traffic congestion is directly impacting the quality of human life. Vehicular congestion is identified by using traffic density metric. CVT based traffic density estimation techniques have high accuracy and real-time congestion detection capabilities which makes it practically feasible to be used in traffic management schemes.

Traffic status is identified in Intent of Vehicles based solutions by using CVT. Novel algorithms are used to classify and process sensory data collected from vehicles to identify congestion levels. IoV systems require support for mobility and location awareness for exchange of time sensitive information over limited bandwidth which makes distributed computing paradigm of Fog computing most suitable architecture for the vehicular WSN. Network traffic and communication latencies can be reduced significantly by deploying Fog Computing based principles in the IoV system which enables the solution to handle larger number of vehicular clients without compromising Quality of Service and Quality of Performance.

## References

[1] J. Barrachina, P. Garrido, M. Fogue, F. Martinez, J. Cano, C. Calafate and P. Manzoni, "A V2I-Based Real-Time Traffic Density Estimation System in Urban Scenarios", Wireless Personal Communications, vol. 83, no. 1, pp. 259-280, Jun. 2015.

[2] C. A. R. L. Brennand, F. D. da Cunha, G. Maia, E. Cerqueira, A. A. F. Loureiro and L. A. Villas, "FOX: A traffic management system of computer-based vehicles FOG," *2016 IEEE Symposium on Computers and Communication (ISCC)*, Messina, 2016, pp. 982-987.

[3] C. A. R. L. Brennand, A. Boukerche, R. Meneguette and L. A. Villas, "A novel urban traffic management mechanism based on FOG," *2017 IEEE Symposium on Computers and Communications (ISCC)*, Heraklion, 2017, pp. 377-382.

[4] T. Darwish and K. Abu Bakar, "Traffic density estimation in vehicular ad hoc networks: A review," *Ad Hoc Networks*, vol. 24, pp. 337–351, Jan 2015.

[5] O. Asmaa, K. Mokhtar and O. Abdelaziz, "Road traffic density estimation using microscopic and macroscopic parameters", *Image and Vision Computing*, vol. 31, no. 11, pp. 887-894, Nov 2013.

[6] J. A. Sanguesa *et al.*, "An infrastructureless approach to estimate vehicular density in urban environments," *Sensors (Switzerland)*, vol. 13, no. 2, pp. 2399–2418, Feb 2013.

[7] J. A. Sanguesa *et al.*, "Sensing traffic density combining V2V and V2I Wireless Communications," *Sensors (Switzerland)*, vol. 15, no. 12, pp. 31794–31810, Dec. 2015.

[8] S. M. Khan, K. C. Dey and M. Chowdhury, "Real-Time Traffic State Estimation with Connected Vehicles," in *IEEE Transactions on Intelligent Transportation Systems*, vol. 18, no. 7, pp. 1687-1699, July 2017.

[9] E. A. Stanciu, I. M. Moise, and L. M. Nemtoi, "Optimization of urban road traffic in Intelligent Transport Systems,", *2012 International Conference on Applied and Theoretical Electricity (ICATE)*, Craiova, 2012, pp. 1–4.

[10] S. M. Bilal, A. U. R. Khan, S. U. Khan, S. A. Madani, B. Nazir, and M. Othman, "Road oriented traffic information system for vehicular ad hoc networks," *Wireess. Personal Commununication*, vol. 77, no. 4, pp. 2497–2515, Feb. 2014.

[11] V. Tyagi, S. Kalyanaraman and R. Krishnapuram, "Vehicular Traffic Density State Estimation Based on Cumulative Road Acoustics," in *IEEE Transactions on Intelligent Transportation Systems*, vol. 13, no. 3, pp. 1156-1166, Sept. 2012.

[12] F. Yang, S. Wang, J. Li, Z. Liu, and Q. Sun, "An overview of Internet of Vehicles," *China Communications*, vol. 11, no. 10, pp. 1–15, Oct. 2014.

[13] A. Boukerche and R. De Grande, "Vehicular cloud computing: Architectures, applications, and mobility", *Computer Networks*, vol. 135, pp. 171-189, Feb. 2018.

[14] H. J. Desirena Lopez, M. Siller and I. Huerta, "Internet of vehicles: Cloud and fog computing approaches," *2017 IEEE International Conference on Service Operations and Logistics, and Informatics (SOLI)*, Bari, 2017, pp. 211-216.

[15] X. Hou, Y. Li, M. Chen, D. Wu, D. Jin and S. Chen, "Vehicular Fog Computing: A Viewpoint of Vehicles as the Infrastructures," in *IEEE Transactions on Vehicular Technology*, vol. 65, no. 6, pp. 3860-3873, June 2016.

[16] C. Huang, R. Lu and K. K. R. Choo, "Vehicular Fog Computing: Architecture, Use Case, and Security and Forensic Challenges," in *IEEE Communications Magazine*, vol. 55, no. 11, pp. 105-111, Nov. 2017.

[17] A. Thakur, R. Malekian, and D. C. Bogatinoska, "Internet of Things Based Solutions for Road Safety and Traffic Management in Intelligent Transportation Systems," *ICT Innovations 2017 Communications in Computer and Information Science*, vol. 778, pp. 47–56, Sept. 2017.

[18] S. Abdelhamid, H. S. Hassanein, and G. Takahara, "Vehicle as a Mobile Sensor," *Procedia Computer Science*, vol. 34, pp. 286–295, Aug. 2014.

[19] J. Barrachina *et al.*, "V2X-d: A vehicular density estimation system that combines V2V and V2I communications," *2013 IFIP Wireless Days (WD)*, Valencia, 2013, pp. 1-6.

[20] R. Bauza and J. Gozalvez, "Traffic congestion detection in large-scale scenarios using vehicle-to-vehicle communications," *Journal of Network and Computer Applications*, vol. 36, no. 5, pp. 1295–1307, Sept. 2013.

[21] G. Agosta *et al.*, "V2I Cooperation for Traffic Management with SafeCop," *2016 Euromicro Conference on Digital System Design (DSD)*, Limassol, 2016, pp. 621-627.

[22] S. Yi, Z. Hao, Z. Qin and Q. Li, "Fog Computing: Platform

and Applications," *2015 Third IEEE Workshop on Hot Topics in Web Systems and Technologies (HotWeb)*, Washington, DC, 2015, pp. 73-78.
[23] R. Mahmud, R. Kotagiri, and R. Buyya, "Fog Computing: A Taxonomy, Survey and Future Directions," *Internet of Things Internet of Everything*, pp. 103–130, Oct. 2017.
[24] B. R. Stojkoska and K. Trivodaliev, "Enabling internet of things for smart homes through fog computing," *2017 25th Telecommunication Forum (TELFOR)*, Belgrade, 2017, pp. 1-4.
[25] M. Wang, J. Wu, G. Li, J. Li, Q. Li and S. Wang, "Toward mobility support for information-centric IoV in smart city using fog computing," *2017 IEEE International Conference on Smart Energy Grid Engineering (SEGE)*, Oshawa, 2017, pp. 357-361.
[26] M. Gerla, "Vehicular Cloud Computing," *2012 The 11th Annual Mediterranean Ad Hoc Networking Workshop (Med-Hoc-Net)*, Ayia Napa, 2012, pp. 152-155.
[27] R. Malekian, N. R. Moloisane, L. Nair, B. T. Maharaj, and U. A. K. Chude-Okonkwo, "Design and Implementation of a Wireless OBD II Fleet Management System," *IEEE Sens. J.*, vol. 17, no. 4, pp. 1154–1164, 2017.
[28] J. Prinsloo and R. Malekian, "Accurate vehicle location system using RFID, an internet of things approach," *Sensors (Switzerland)*, vol. 16, no. 6, pp. 1–51, 2016.
[29] A. Aliyu, A. H. Abdullah, O. Kaiwartya, Y. Cao, M. J. Usman, S. Kumar, D. K. Lobiyal, and R. S. Raw, "Cloud Computing in VANETs: Architecture, Taxonomy, and Challenges," *IETE Technical Review*, vol. 4602, pp. 1–25, Aug. 2017.
[30] A. Rayamajhi *et al.*, "ThinGs In a Fog: System Illustration with Connected Vehicles," *2017 IEEE 85th Vehicular Technology Conference (VTC Spring)*, Sydney, NSW, 2017, pp. 1-6.
[31] M. Whaiduzzaman, M. Sookhak, A. Gani, and R. Buyya, "A survey on vehicular cloud computing," *Journal of Network and Computer Applications.*, vol. 40, no. 1, pp. 325–344, Aug. 2013.
[32] B. R. Stojkoska, K. Trivodaliev, and D. Davcev, "Internet of things framework for home care systems", *Wireless Communications and Mobile Computing.*, vol. 2017, March 2017.
[33] A. V. Dastjerdi and R. Buyya, *Internet of things: principles and paradigms*. Cambridge, MA: Morgan Kaufmann, 2016.
[34] E. Baccarelli, P. G. V. Naranjo, M. Scarpiniti, M. Shojafar, and J. H. Abawajy, "Fog of Everything: Energy-Efficient Networked Computing Architectures, Research Challenges, and a Case Study," *IEEE Access*, vol. 5, pp. 9882–9910, 2017.
[35] W.-Y. Lin, M.-W. Li, K.-C. Lan, and C.-H. Hsu, "A Comparison of 802.11a and 802.11p for V-to-I Communication: A Measurement Study," in *Quality, Reliability, Security and Robustness in Heterogeneous Networks: 7th International Conference on Heterogeneous Networking for Quality, Reliability, Security and Robustness, QShine 2010, and Dedicated Short Range Communications Workshop, DSRC 2010*, X. Zhang and D. Qiao, Eds. Berlin, Heidelberg: Springer Berlin Heidelberg, 2012, pp. 559–570.
[36] A. M. de Souza, C. A. Brennand, R. S. Yokoyama, E. A. Donato, E. R. Madeira, and L. A. Villas, "Traffic management systems: A classification, review, challenges, and future perspectives," *Int. J. Distrib. Sens. Networks*, vol. 13, no. 4, p. 1-14, 2017.
[37] R. Zhang, Y. Ma, Z. Li, R. Malekian, and M. A. Sotelo, "Energy Dissipation Based Longitudinal and Lateral Coupling Control for Intelligent Vehicles," *IEEE Intell. Transp. Syst. Mag.*, vol. 10, no. 2, pp. 121–133, 2018.
[38] S. Chun, S. Shin, S. Seo, S. Eom, J. Jung and K. H. Lee, "A Pub/Sub-Based Fog Computing Architecture for Internet-of-Vehicles," *2016 IEEE International Conference on Cloud Computing Technology and Science (CloudCom)*, Luxembourg City, 2016, pp. 90-93.
[39] M. Mahbadi, M. M. Manohara Pai, S. Mallissery and R. M. Pai, "Cloud-enabled vehicular congestion estimation: An ITS application," *2016 IEEE Canadian Conference on Electrical and Computer Engineering (CCECE)*, Vancouver, 2016, pp. 1-4.
[40] S. Kaklij, "Mining GPS Data for Traffic Congestion Detection and Prediction", *International Journal of Science and Research (IJSR)*, vol. 4, no. 9, pp. 976-980, Sept. 2015.
[41] P. Gora, "Simulation-based traffic management system for connected and autonomous vehicles", *Road Vehicle Automation*, vol. 4, pp. 257-266, 2018.
[42] Y. Xu, Y. Wu, J. Xu, T. Liu, J. Wang and A. Lin, "An efficient detection scheme for urban traffic condition using volunteer probes," *2014 20th IEEE International Conference on Parallel and Distributed Systems (ICPADS)*, Hsinchu, 2014, pp. 768-773.
[43] H. El-Sayed, G. Thandavarayan, S. Sankar and I. Mahmood, "An infrastructure based congestion detection and avoidance scheme for VANETs," *2017 13th International Wireless Communications and Mobile Computing Conference (IWCMC)*, Valencia, 2017, pp. 1035-1040.
[44] H. El-Sayed and G. Thandavarayan, "Urban Area Congestion Detection and Propagation Using Histogram Model," *2016 IEEE 84th Vehicular Technology Conference (VTC-Fall)*, Montreal, 2016, pp. 1-6.
[45] M. Milojevic and V. Rakocevic, "Short paper: Distributed vehicular traffic congestion detection algorithm for urban environments," *2013 IEEE Vehicular Networking Conference*, Boston, MA, 2013, pp. 182-185.
[46] M. Artimy, "Local Density Estimation and Dynamic Transmission-Range Assignment in Vehicular Ad Hoc Networks," in *IEEE Transactions on Intelligent Transportation Systems*, vol. 8, no. 3, pp. 400-412, Sept. 2007.
[47] Y. Xu, J. Wang, T. Liu, W. Yu and J. Xu, "Detecting Urban Road Condition and Disseminating Traffic Information by VANETs," *2015 IEEE 12th Intl Conf on Ubiquitous Intelligence and Computing and 2015 IEEE 12th Intl Conf on Autonomic and Trusted Computing and 2015 IEEE 15th Intl Conf on Scalable Computing and Communications and Its Associated Workshops (UIC-ATC-ScalCom)*, Beijing, 2015, pp. 93-98.
[48] L. Zilu and Y. Wakahara, "City Traffic Prediction based on Real-Time Traffic Information for Intelligent Transport Systems," ITS Telecommun. (ITST), 2013 13th Int. Conf., pp. 378–383, 2013.
[49] Z. Wang, N. Ye, F. Xiao, and R. Wang, "TrackT: Accurate tracking of RFID tags with mm-level accuracy using first-order taylor series approximation," *Ad Hoc Networks*, vol. 53, pp. 132–144, 2016.
[50] Y. Ma, Z. Li, , R. Zhang, X. Song and M. A. Sotelo, "Hierarchical Fuzzy Logic-Based Variable Structure Control for Vehicles Platooning," in *IEEE Transactions on Intelligent Transportation Systems*. pp. 1–12, 2018.